# Narrow linewidth quantum emitters in electron-beam shaped polymer

Carlotta Ciancico[1,*], Kevin G. Schädler[1,*], Sofia Pazzagli[2], Maja Colautti[2,3], Pietro Lombardi[2], Johann Osmond[1], Camilla Dore[4], Agustín Mihi[4], Anna P. Ovvyan[5], Wolfram H. P. Pernice[5], E. Berretti[6], A. Lavacchi[6], Costanza Toninelli[2,7], Frank Koppens[1,8], Antoine Reserbat-Plantey[1]

[1]ICFO – Institut de Ciencies Fotoniques, The Barcelona Institute of Science and Technology, 08860 Castelldefels (Barcelona), Spain

[2]LENS and CNR-INO, Via Nello Carrara 1, 50019 Sesto Fiorentino (FI), Italy

[3]Dipartimento di Fisica ed Astronomia, Università di Firenze, Via Sansone 1, 50019 Sesto Fiorentino (FI), Italy

[4]ICMAB-CSIC – Institut de Ciència de Materials de Barcelona, Campus de la UAB, 08193 Bellaterra (Barcelona), Spain

[5]Physikalisches Institut, Westfälische Wilhelms, Universität Münster, Heisenbergstrasse 11, 48149 Münster, Germany

[6]Institute for Chemistry of OrganoMetallic Compounds (ICCOM – CNR), Via Madonna del Piano, 50019 Sesto Fiorentino (FI), Italy

[7]QSTAR, Largo Fermi 2, 50125 Firenze, Italy

[8]ICREA – Institució Catalana de Recerça i Estudis Avancats, 08010 Barcelona, Spain

\* = equal contributions
Corresponding authors: frank.koppens@icfo.eu – antoine.reserbat-plantey@icfo.eu

*Abstract*

Solid-state single photon sources (SPSs) with narrow linewidth play an important role in many leading quantum technologies. Within the wide range of SPSs studied to date, single fluorescent molecules hosted in organic crystals stand out as bright, photostable SPSs with lifetime-limited optical resonance at cryogenic temperatures. Furthermore, recent results have demonstrated that photostability and narrow linewidths are still observed from single molecules hosted in a nanocrystalline environment, which paves the way for their integration with photonic circuitry. Polymers offer a compatible matrix for embedding nanocrystals and provide a versatile yet low-cost approach for making nanophotonic structures on chip that guide light and enhance coupling to nanoscale emitters. Here, we present a deterministic nanostructuring technique based on electron-beam lithography for shaping polymers with embedded single molecules. Our approach provides a direct mean of structuring the nanoscale environment of narrow linewidth emitters while preserving their emission properties.

Ideal SPSs are isolated quantum systems displaying stable emission within a narrow linewidth[1], which is fundamental for quantum technology applications such as entanglement and single photon interference[2,3]. At the same time, harnessing such SPSs requires controlling their emission, for instance by integration with nanophotonic structures at the wavelength scale (e.g. photonic crystal cavities[4] or waveguides[5,6]) and below (plasmonic structures[7]). However, the integration process typically deteriorates desirable SPS properties due to i) the nanofabrication steps involved and ii) enhanced coupling to environmental fluctuations such as surface states and charge puddles that lead to spectral broadening, jittering, wandering or photobleaching[8]. Quantum emitters embedded in a nanoscale environment[9] convenient for nanophotonic integration suffer particularly strongly from this effect. For instance, epitaxial InAs/GaAs quantum dots in GaAs nanowires[10–12] or strain-mediated quantum dots in single layers[13,14] of $WSe_2$ are recent examples of SPSs that can be deterministically positioned, but the observation of narrow optical resonances from such emitters remains challenging. Polymers offer an interesting platform for structuring the environment of SPSs at the nanoscale in order to couple them to photonic structures[15,16] on chip. While lithographic techniques have already shown impressive results for on-chip integration of nitrogen-vacancy centres[17] and quantum dots[18], the outstanding properties of organic single molecules as quantum emitters and their compatibility with polymers motivate the quest for novel approaches based on electron-beam lithography to achieve nanophotonic integration while preserving their emission properties.

Our work is based on dibenzoterrylene (DBT) molecules hosted in anthracene nanocrystals fabricated by reprecipitation method[9]. We modified this technique to make a suspension of nanocrystals in a polymer instead of water by injecting a solution of DBT (10 µM) and anthracene dissolved in acetone into a sonicating polyvinyl alcohol (PVA) water solution (5%/weight). We have also developed a similar protocol with another polymer, hydroxypropyl cellulose (HPC), as will be discussed later. This suspension of nanocrystals in PVA - termed "doped-PVA" in the following - can be spin-cast on a standard $Si^{++}/SiO_2$ substrate (Fig. 1a). Reference measurements on water-based suspensions show that nanocrystals in such films display typical dimensions ranging from 50 to 500 nm, with a distribution peaked at 230 nm (cf. figure S1). PVA is a water-based polymer which combines two advantageous properties for nanophotonic integration of quantum emitters: i) it is insoluble in most organic solvents used in nanofabrication and thus protects encapsulated nanocrystals[19] and ii) can be deterministically structured by electron-beam cross-linking[20,21], using water as a developer. Together, these two properties enable the fabrication of three-dimensional structures (Fig. 1b-c) which can host and protect quantum emitters. Importantly, this e-beam patterning enables extreme lateral nanostructuring. The height difference due to the cross-linking process is appreciable in the change of colour in the exposed structures. In our study, we have realized structures ranging from micron sizes down to 240 nm wide pillars (Fig. 1d). Furthermore, complex structures with non-uniform height such as mesas with shallow holes (Fig. 1c) and lens-like (Fig. 1e) structures can be fabricated by locally varying the electron beam dose, which controls the height of cross-linked structures. For instance, Fig. 1e shows the atomic force microscopy (AFM) profile of concentric rings obtained by decreasing the e-beam dose from the centre to the edges (see also Fig. S2). Measurements on structures fabricated over a large range of doses (1-10 $mC/cm^2$) show that our technique provides continuous structure height control up to 80% of the initial PVA film thickness, independent of the initial thickness $t_i$ (Fig. 1f). These measurements show that electron-beam lithography of PVA enables nanoscale 3D shaping quantum emitter environment.

To evaluate the fluorescence properties of the DBT molecules in the structure PVA, we first characterize the emission properties at room temperature. To this end, we use a custom microscope with wide field LED illumination at 730nm which off-resonantly excites DBT molecule ensembles, as reported previously[9,22]. DBT emission is then imaged using a camera (Fig. 1i). Reflection and fluorescence images of an exposed square area after development at ambient conditions are shown in Fig. 1g-h. Due to the cross-linking process, exposed areas have a different contrast to non-exposed ones (Fig. 1g). An emission map (Fig. 1h) reveals bright spots randomly dispersed on the sample in both exposed and non-exposed areas, which indicates that electron-beam irradiation does not inhibit the emission of DBT molecules.

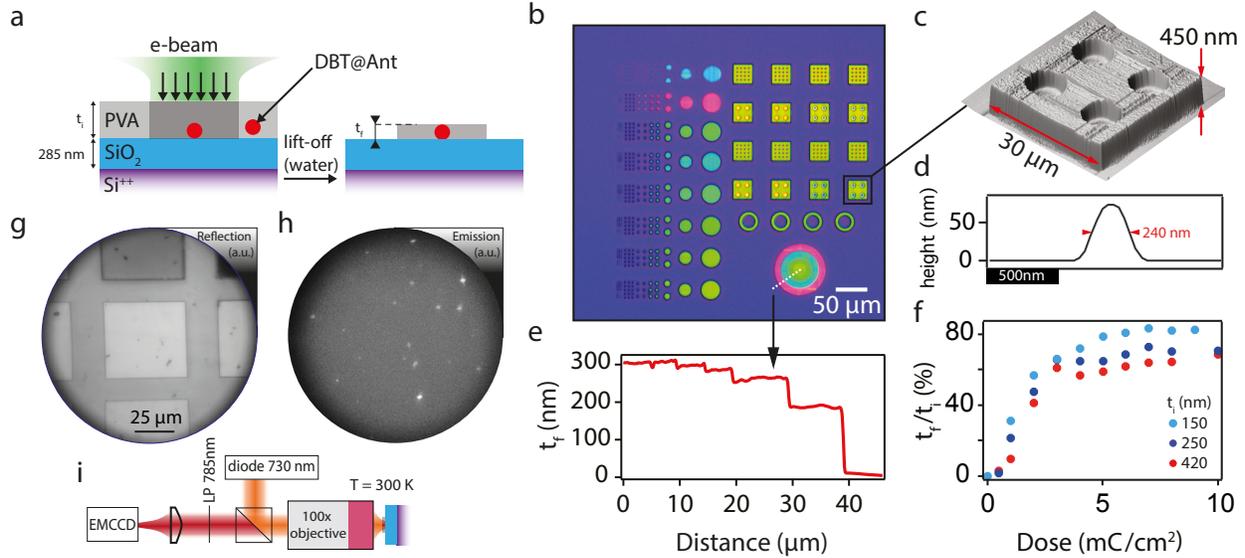

**Figure 1: Nanostructuring PVA thin films hosting anthracene nanocrystals by electron beam lithography (EBL). a:** DBT molecules are contained inside anthracene nanocrystals (labelled as DBT@Ant), randomly dispersed in a PVA layer. Zones that are non-exposed to electron-beam are lifted-off in water. Optical (**b**) and atomic force (**c**) micrographs of nano-structured PVA. **d:** AFM profile of a lithographed PVA nano-pillar (not shown in **b**). **e:** Height profile of nano-structured PVA measured by atomic force microscopy (dashed line in **b**). **f:** Ratio of remaining over initial PVA thickness as a function of EBL dose for three different initial film thicknesses. **g:** Optical reflection at 730 nm of patterned PVA squares and corresponding room temperature DBT emission (**h**). Squares of different contrasts correspond to exposed areas with various e-beam doses. **i:** Sketch of the room temperature wide field fluorescence imaging setup.

Next, we study the optical emission linewidth of DBT at low temperature with a custom cryogenic confocal scanning microscope (Fig. 2a) with single molecule fluorescence excitation spectroscopy[23]. Samples are cooled to 2.7 K and illuminated with a tuneable laser at 785 nm that excites the zero-phonon line (00ZPL) of DBT molecules. Intensity is kept below 1 W/cm$^2$ to avoid emission saturation (cf. figure S3). By sweeping the laser frequency and collecting red-shifted DBT fluorescence, the excitation spectrum of single molecules is detected. For instance, Fig. 2b (bottom) shows an excitation spectrum of an ensemble of DBT molecules in a single nanocrystal contained in nanopatterned PVA resist irradiated by an electron dose (1 mC/cm$^2$). A typical peak within the inhomogeneously broadened ensemble (top panel of Fig. 2b) displays a linewidth of $\Gamma = 2\pi \cdot 134 \pm 5$ MHz. The photon correlation function measured in Hanbury-Brown and Twiss configuration under resonant fluorescence excitation of a single peak (Fig. 2c) shows strong antibunching with g$^2$(0) = 0.08 $\pm$ 0.02. This observation confirms that the spectral peaks observed correspond to the emission from single molecules and that single photon emission is preserved upon electron irradiation.

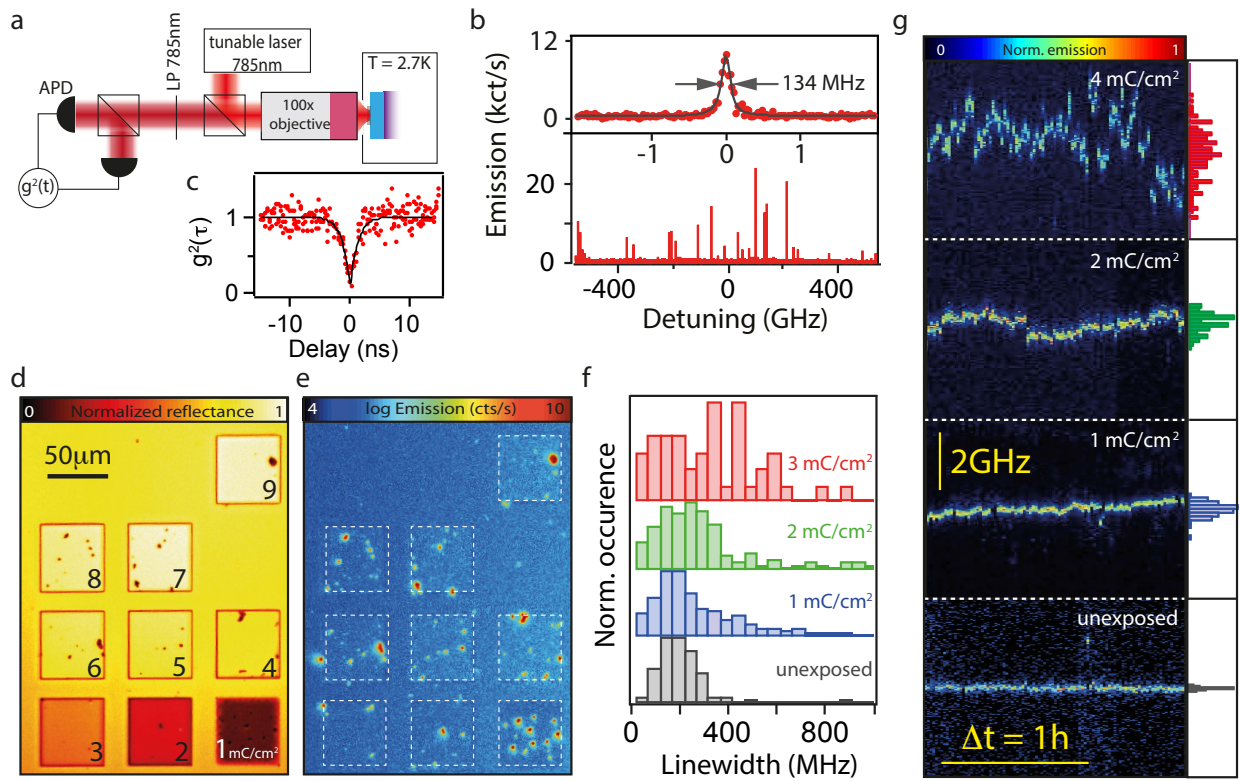

**Figure 2: Narrow quantum emitters integrated in nanostructures. a:** Sketch of the optical cryogenic setup to detect emission from single molecules. **b:** Fluorescence excitation spectrum of an ensemble (bottom) of single molecules (top: zoomed spectrum) in a nanocrystal at 2.7 K that have been exposed to e-beam (dose = 1 mC/cm$^2$). **c:** Anti-bunching measurement for resonant excitation of a single peak. The solid line is a fit to the data using a second-order correlation function. **d:** Optical micrograph of patterned squares of PVA containing nanocrystals exposed to different electron doses (from 1 to 9 mC/cm$^2$). **e:** DBT fluorescence map (integrated over all laser detuning) of the sample shown in **d**. **f:** Emission peak linewidth distribution for DBT subjected to various electron beam doses. **g:** Time traces of single DBT molecules emission for unexposed, 1, 2 and 4 mC/cm$^2$ e-beam doses (from bottom to top). Normalized histograms of central peak position are shown on the right to quantify the emitter's stability in time.

Having observed single photon emission from electron-irradiated single DBT molecules, we now investigate the impact of electron dose on the emission linewidth and time stability of the molecules. First, we prepare an array of mesas irradiated with increasing electron dose ranging from 1 to 9 mC/cm$^2$ (Fig. 2d-e). In all exposed areas, we observe bright emission spots over the full laser detuning (Fig. 2e) corresponding to nanocrystal positions easily identified in the reflection image (Fig. 2d). DBT emission is only observed in the exposed areas as nanocrystals in non-exposed ones are washed away during the development process.

First, we measure the spectra of a total of 808 single DBT emission peaks subjected to different e-beam doses, to obtain a linewidth histogram as a function of the e-beam dose (Fig. 2f). The emission linewidths are extracted from spectra measured via scanning laser spectroscopy[24] on anthracene nanocrystals within the mesa, with an averaging time of 0.01s/point and a resolution of 0.02 GHz.

Our results show that low dose exposure (1 mC/cm$^2$) does not significantly broaden the DBT emission line in comparison to a control experiment (unexposed sample). With gradually increasing dose, larger emission linewidths are observed which broaden the distribution. We attribute this result to a modification of the molecule's environment due to detrimental effects of the electron radiation such as electrostatic charging, thermoplastic deformation, hydrocarbon contamination and knock-on mechanisms[25,26] that affect the crystallinity of the host nanocrystal. Interestingly, a fraction of narrow linewidth ($< 2\pi \cdot 200$ MHz) peaks remains for each e-beam dose studied. This is indicative of a dominant surface effect of the e-beam on the nanocrystal. This can be understood assuming a random distribution of molecules within the nanocrystal: molecules closer to the surface are strongly affected by any surface reconstruction[27], electrostatic puddles[28] or amorphous carbon contaminants[29] which can lead to a broadening of emission lines. Conversely, molecules closer to the centre of the nanocrystal are less perturbed and exhibit narrow linewidth similar to non-exposed samples.

To determine the influence of electron beam irradiation on the spectral stability of the molecules, we monitor the behaviour of single molecule emission spectra in time for various doses (Fig. 2g). At lower dose, we observe stable lines over a time scale of hours, similar to the unexposed control sample. In contrast, measurements on emitters irradiated with higher doses reveal emission spectral jittering which can be due to the presence of fluctuating charge puddles created by e-beam exposure. Interestingly, it appears that even high electron irradiation doses do not induce significant emission blinking.

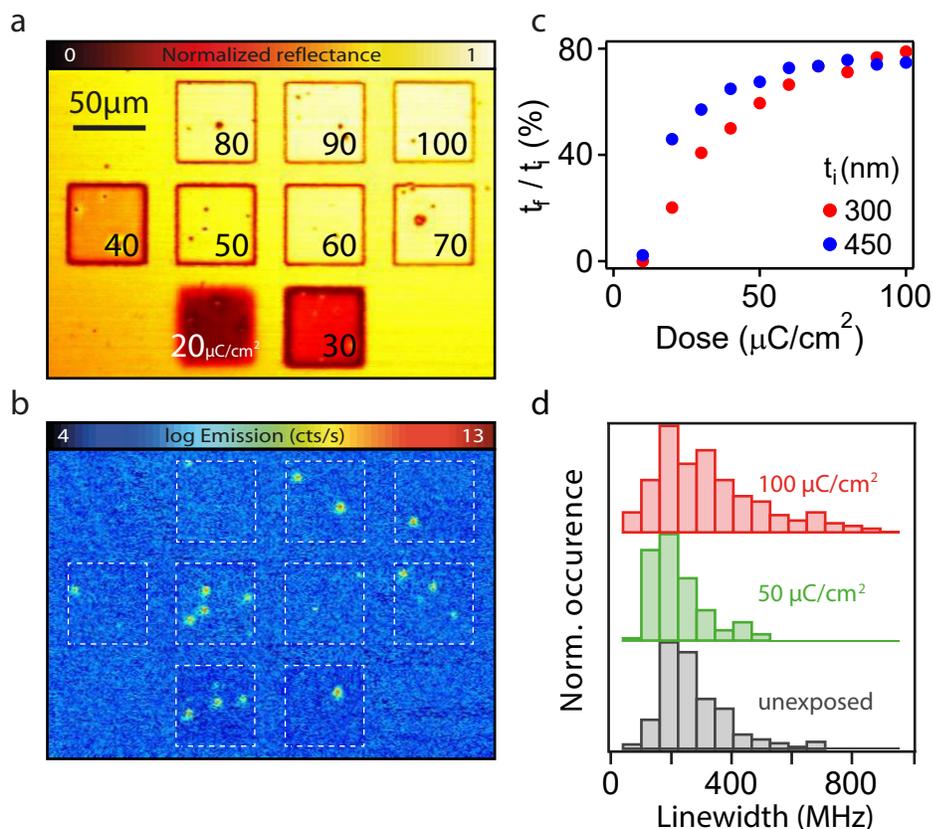

**Figure 3: Low-dose nanolithography on HPC thin films doped with anthracene nanocrystals. a:** Optical micrograph of patterned squares of HPC containing nanocrystals of DBT@Ant exposed to different electron doses (from 20 to 100 µC/cm²). **b:** DBT fluorescence map (integrated over all laser detuning) of the sample shown in **a**. **c:** Ratio of remaining over initial HPC thickness as a function of EBL dose for two different initial film thicknesses. **d:** Emission peak linewidth distribution for DBT molecules subjected to various electron beam doses.

To highlight the versatility of our method and reduce even more the e-beam dose needed to create the nanostructures, we use another water-based polymer, hydroxypropyl cellulose (HPC)[30]. Upon irradiation of an HPC film doped with DBT molecules with increasing dose ranging from 20 to 100 µC/cm², two orders of magnitude lower than for PVA, we pattern an array of mesas containing fluorescent nanocrystals (Fig. 3a-b). From this, we can control the height of up to 80% of the initial film thickness, independent of the initial thickness $t_i$ (Fig. 3c), similarly to the case of PVA. At such low doses, detrimental effects due to electron radiation are drastically reduced, as shown in Fig. 3d. At the maximum dose (100 µC/cm$^2$), we still observe that the most probable linewidths within the ensemble are similar to the unexposed case. Thus, EBL on HPC enables a more extensive height control of the doped polymer layer while exposing the molecules to a lower e-beam dose.

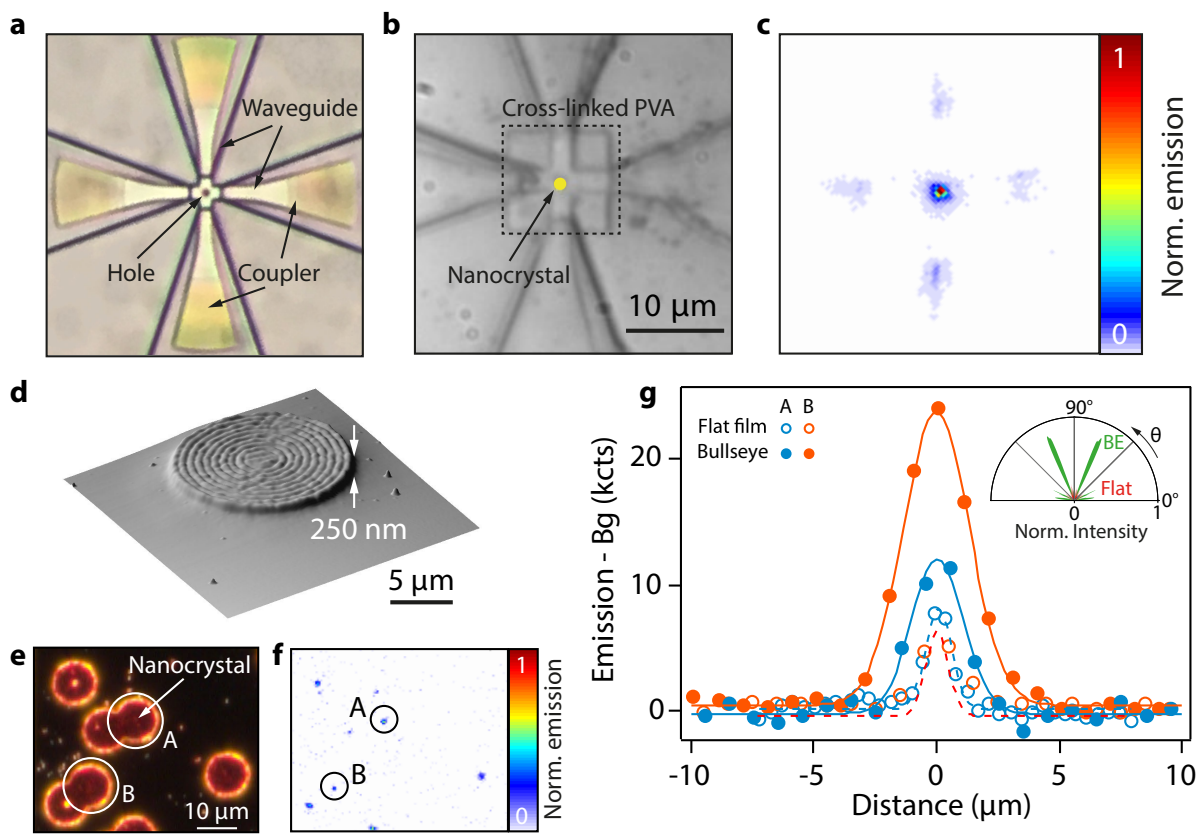

**Figure 4: Hybrid nanophotonic devices with integrated DBT molecules. a:** Crossed silicon nitride waveguide (WG) with a circular aperture of 1 µm at the intersection. **b:** Reflection image of a WG as illustrated in (**a**) after deposition of DBT@Ant nanocrystals and EBL patterning of a protective PVA layer. **c:** Fluorescence map collected with an EMCCD camera under off-resonant confocal excitation at the nanocrystal position for the WG shown in (**b**). The out-coupled light at the four grating couplers is clearly visible. **d:** Atomic force microscopy (AFM) image of a bullseye structure exposed in HPC. The coupler is made of 10 rings and has a period of 550 nm and gap of 165 nm. **e.** Dark field reflection image of bullseye gratings aligned on the anthracene nanocrystal positions. **f.** Fluorescence image from the sample in (**e**). **g.** Comparison of fluorescence at RT for two molecules in a cellulose flat film and after integration within the bullseye antennas shown in (**e**). Enhancement of emission up to a factor 2 is observed, as confirmed by the simulation of the radiation pattern (inset).

We now present two applications of our method showing integration of narrow-linewidth emitters in nanophotonics devices. Fig. 4a shows a silicon nitride (SiN) crossed waveguide (WG) on a SiO$_2$ substrate. The WG has a 1 µm circular aperture at the intersection of the four branches. First, we perform micro-infiltration of DBT@Ant nanocrystals into the aperture at the centre of the WG (Fig. 4b). The micro-infiltration set-up (Eppendorf Femtojet) consists of a micropipette (Eppendorf Femtotips) with external diameter of about 2 µm and inner diameter of 0.5 µm, held on a 3D micrometric stage for fine movement. The aqueous nanocrystals suspension is injected into the micropipette, upon filtering with a 450 µm pore-size filter (Sartorius Minisart) to get rid of eventual clusters and avoid obstruction. By monitoring the process with an optical microscope, the pipette tip is then approached to the region of interest until a micro-drop of suspension is deposited via surface adhesion. After water evaporation, nanocrystals are positioned with high precision and good success rate (∼ 1/3 of deposited NCs are at the centre of the crossed waveguide structure). Then, PVA squares are patterned by electron-beam lithography at the intersection of the crossed WG. As a result, EBL shaping allows to prevent crystal sublimation at ambient condition and provides a clean surface, which is crucial to achieve efficient light coupling into the waveguide. Off-resonant confocal excitation at 767 nm triggers emission from the molecules which partially couples into the waveguide and is guided throughout the four grating out-couplers (Fig. 4c). By evaluating the relative intensity of the guided emission, coupling of the molecule to the structure is estimated to be about 10%. This value is the result of the integration of the emission of many molecules in the nanocrystal, each with unequal coupling efficiency owing to the different position and orientation, and is therefore an underestimation. These results show the successful insertion of the nanocrystal at the centre of the WG and the potential of our technique to couple quantum emitters to standard photonics structures.

The polymer itself can also provide the platform to design nanophotonic structures aligned on the nanocrystals positions. A circular "bullseye" grating with a period of 550 nm and gap of 165 nm is modelled and exposed on a layer of HPC on a gold substrate. This geometry serves to maximize the out-coupling of an optical excitation from the emitter into the far-field (Fig. 4d). The fabrication starts with the measurement of the fluorescence image of DBT molecules in the HPC layer, deposited on a substrate with pre-patterned alignment markers. The bullseye design is then aligned on the extracted positions with respect to the markers (Fig. 4e-f). By comparing the emission from nanocrystal positions A and B before and after integration into the bullseye antenna, we observe emission enhancement of up to a factor 2 (Fig. 4g). The increased signal in the presence of the bullseye structure can be interpreted as an increased collection efficiency, as confirmed by numerical simulations of the emission pattern of a structure with the same nominal parameters (see inset of fig.4). It is worth noticing that emission enhancement is mainly observed from nanocrystals that are not clearly visible in the dark field reflection image (Fig. 4e). This is due to the fact that nanocrystals more deeply embedded in the HPC layer couple more efficiently to the bullseye antenna. More details on the simulations are presented in figure S4.

Our observation of narrow linewidth single photon emission from electron-irradiated DBT molecules highlights the potential of electron beam patterning for single-molecule-based quantum photonic devices. We envision different geometries such as waveguides[6], ring resonators, or "bullseye" antennae[31] and pillar structures for efficient out-of-plane light extraction and emission control. Combining PVA with additives[32] for increasing its nominally low refractive index ($n_{PVA} \sim 1.5$) would enhance light confinement on substrates of similar refractive index. Further, solid-state quantum emitters properties are strongly affected by their near-field environment (plasmon-molecule coupling[33,34], Casimir[35], energy transfer[22,36]). Our patterning technique opens new possibilities exploring such near-field effects by shaping the emitter's environment with high precision, as well as combining it with other materials (conductive films, 2D materials). Such hybrid systems enable manipulation of the emitter properties at the scale of a single molecule, which would have a strong impact for quantum nanophotonics[37] and optomechanics[35,38].

*Supporting Information*
Details on anthracene nanocrystal formation, PVA structures, single molecules characterization and simulations of bullseye antennae.

*Acknowledgements*

We would like to thank Josep Canet Ferrer and Vittoria Finazzi. This project has received funding from the EraNET Cofund Initiatives QuantERA under the European Union's Horizon 2020 research and innovation programme grant agreement n° 731473 (project acronyme: ORQUID). C.C. acknowledges financial support by the ICFOstepstone - PhD Programme for Early-Stage Researchers in Photonics, funded by the Marie Skłodowska-Curie Co-funding of regional, national and international programmes (GA665884) of the European Commission. C.D. acknowledges financial support by the European Union's Horizon 2020 research and innovation programme (ERC grant no. StG637116). We also acknowledge financial support from the Spanish Ministry of Economy and Competitiveness (MINECO), through the "Severo Ochoa" Programme for Centres of Excellence in R&D (SEV-2015-0522 and SEV-2015-0496), support by Fundaciò Cellex Barcelona, Generalitat de Catalunya through the CERCA program. This work received funding from the European Union's Horizon 2020 research and innovation program Quantum Flagship (grant no. 820378).